\documentclass[12pt]{article}

\usepackage{graphics}
\usepackage{amsmath}
\def\ni{\noindent}
\def\ph{{\phantom{...}}}
\def\={\phantom{..} = \phantom{..}}

\def\+{\phantom{..} + \phantom{..}}
\def\>{\phantom{..} > \phantom{..}}
\def\<{\phantom{..} < \phantom{..}}
\def\-{\phantom{..} - \phantom{..}}

\def\vs{{\em vs.}}

\def\bq{\begin{quote}}
\def\eq{\end{quote}}
\def\be{\begin{equation}}
\def\ee{\end{equation}}
\def\bar{\begin{eqnarray}}
\def\ear{\end{eqnarray}}
\def\no{\nonumber}

\def\Sch{Schr{\"o}dinger}

\def\sdoic{sensitive dependence on initial conditions}
\def\TDotS{That Dot on the Screen}

\def\skN{\sum_{k=1}^N}
\def\slN{\sum_{l=1}^N}

\title{\bf Nonlinear Nonlocal: Comparing A. O. Barut's Theory to Mine\\[0.5in]
 with special emphasis on That Dot on the Screen\\[3in]}

\author{W. David Wick\footnote{email: wdavid.wick@gmail.com}}

\begin{document}
\maketitle
\pagebreak

\section*{Abstract}
In the 1980's and 90's, A. O Barut and colleagues developed a 
nonperturbative approach to electrodynamics eschewing so-called ``second-quantization".
Based on incorporation of self-energy terms, the resulting nonlinear and nonlocal 
theory explained many well-known phenomena of atomic and radiation physics. 
In 2017, this author introduced a nonlinear, nonlocal theory with the intent of
resolving the Measurement Problem. Barut also suggested that his theory resolved such
paradoxes. Here I compare the two theories with special attention to \TDotS.

\section{Introduction: Barut's Nonlinear Nonlocal\\ (and Non-quantum) Electrodynamics \label{introsection}}

Beginning in the early 1980's, A. O. Barut and colleagues pursued a program of
explaining electrodynamics entirely within wavefunction theory. The model they chose
was the combined Maxwell-Dirac equations, which can be written in terms of fields:

\be
\psi_1(x), \psi_2(x),...,\psi_N(x) \ph \hbox{and} \ph A_{\mu}(x),
\ee

\ni which they used to represent classical (``first-quantized", ``c-valued",
meaning real-or-complex valued) Dirac (4-component) wavefunctions representing $N$ electrons
and the 4-potentials of Maxwell's electromagnetic field. All of these fields
are treated as elements of reality, not stand-ins for some other entities.
The dynamical equations are:

\bar
  \left(\,\gamma^{\mu}\,i\,\partial_{\mu} - m_k\,\right)\,\psi_k &\=& 
e\,\gamma^{\mu}\,\psi_k\,A_{\mu};\\
  F^{\mu,\nu}_{, \nu} &\=& - e\,\skN\,\psi^*_k\,\gamma^{\mu}\,\psi_k \label{Feq} \\
\no  &\=& \skN\,j^{\mu}_k \= j^{\mu}.\\
&&
\ear

\ni Here the field $F^{\mu,\nu}$ (its components are the electric and magnetic fields) is given by:

\be
F^{\mu,\nu} \= A^{\mu,\nu} - A^{\nu,\mu},
\ee

\ni and the $j^{\mu}_k$ are the Dirac charge-currents assocated with the Dirac spinors,
with $j^{\mu}$ the total current. (For a short account, see \cite{Barut1992}.)

Among other things, this set-up solves a problem that had vexed Maxwell: what is the source for his
electric and magnetic fields? (The point electron was not ``discovered", or proposed, until
after Maxwell's death in 1879. Even if it had been, it is a paradoxical source, as the total
energy in the field of a point electric charge is infinite.) Presumably, a continuous
current of charge would have been most satisfactory.
 
Solving (\ref{Feq}) using the Green's function:

\be
A_{\mu}(x) \= \int\,dy\,D(x-y)\,j_{\mu}(y),
\ee

\ni and using this solution gave the action:

\be
\skN\,\left(\,\gamma^{\mu}\,i\,\partial_{\mu} - m_k\,\right)\,\psi_k \- 
(1/2)\,\skN\slN\,\int\,dx\,\int\,dy\,j^k_{\mu}(y)\,D(x - y)\,j^{\mu}_l(x).\label{Baruteq}
\ee

\ni The ``self-energy" is contained in the terms with $k = l$ and is the novel ingredient,
not appearing in conventional QM or QED. Barut remarks:

\begin{quote}

Not surprisingly, it was \Sch\ who first formulated the self-consistent Maxwell and 
matter field equations ... \Sch\ however calculated only the static part of the self-energy
and obtained unacceptibly large self-energies.\footnote{\cite{Barut1991}, section 3.3}

\end{quote}

The second term, which we can regard as a form of potential energy, is of particular
interest to this author as it is nonlinear, nonlocal, and scales as $N^2$. 

It is important to note that this theory is fully defined, 
relativistically-invariant,\footnote{Meaning special-relativity-invariant. 
One can argue that Dirac's theory
of the electron was not relativistically-invariant, because the transformation rule
between frames involves a non-unitary matrix. However, this objection derives from
Wigner's Law about implementing symmetries in quantum theories with the usual probabilistic
interpretation by unitary or anti-unitary
transformations. Barut's theory is not a quantum theory.} and stated in a closed
system of equations. Reflecting on the history of quantum theory, it is fascinating
to note that the following phrases, mathematical machinery, and concepts
are unneeded:

\begin{description}

\item Second quantization 

\item Creation or annihilation operators

\item Perturbation series

\item Series convergence or divergence

\item Feynman diagrams

\item Probabilities

\item Quantum

\item Jumps

\item Particles

\end{description}

I am serious about that last entry. Although Barut and colleagues often refer to particles,
that is due to their respect for the history of the subject 
and their understandable desire to make the material seem familiar. 
One could easily
write a book or give a lecture on their theory never mentioning the ``electron" or
the ``photon".

Here is a partial list of topics this school addressed, often claiming results
as good or better than obtained by the conventional Quantum Electrodynamics (QED) 
party of Feynman, Schwinger, Dyson {\em et al.}:

\begin{description}

\item Spectral lines

\item Spontaneous emission

\item Lamb shift

\item Anomalous magnetic moment

\item Antiparticles

\item Planck's distribution for blackbody radiation

\end{description}

Particularly interesting for me is the first entry: spectral lines in 
emission/absorption spectra of atoms. 
I recall from graduate school that the conventional explanation went like this:

\begin{quote}
Atoms exist in discrete energy levels. Spectral lines are explained by the emission or absorption
of a photon having energy, according to Einstein's formula, $h\,\nu$ where $\nu$ is the photon's
frequency.\footnote{I recall wondering, if the photon is Einstein's light-quantum, how it could
have a certain frequency, as Einstein clearly believed that he was describing a corpuscle,
as Newton had postulated in his {\em Optics}. If a light-quantum is a 
wavepacket localized in space, it must have
a spread of frequencies. If it is not, but some kind of bullet, how can it be said to have
a frequency?} This energy exactly equals the energy difference between two
atomic energy levels.\footnote{About this
language appearing in textbooks and many, many articles in atomic physics: it was always 
imposed ideology.
No matter what flavor of quantum theory you support, the equations never described populations
of atoms in various states jumping between them while emitting or absorbing photons. 
For instance, no such
theory can tell us how long it is between jumps. And the ``photon number"
of QED is an integer appearing
in a linear oscillation theory. The photon is not localized in space.}  

\end{quote}

Contrast that with Barut's account, \cite{Barut1992}:

\begin{quote}
With self-energy the system always has continuous spectrum, with a stable ground state.
Instead of a discrete spectrum, we get spectral concentrations or resonances\footnote{\Sch\
also conjectured that the relation: $E_k - E_j = h\,\nu$ described a resonance
in the atom, rather than energy conservation in a collision, \cite{Sch1952}.}  at certain
energy values. That is also what is observed experimentally, so the discreteness of
quantum theory is an idealization. Nor is it necessary to assume discrete photons.

\end{quote}

\pagebreak

\section{The Author's Nonlinear, Nonlocal Wavefunction Theory\label{WFEsection}}

In 2017, the present author proposed altering \Sch's\ equation from 1926 by incorporating
nonlinear terms in the Hamiltonian. The goal was to resolve the Measurement Problem (MP). 
I'll briefly recap the problem next.

The MP originates in the famous ``cat paper" written by \Sch\ in 1935, \cite{Schcatpaper},
 although it is implicit in
von Neumann's book of 1932, \cite{vNbook}. The context is measurement situations, which necessarily
give rise to ``cats". A ``cat" in physics is an object whose center-of-mass (COM)
has dispersion, as calculated from the wavefunction, larger than the object's size.
Because of the Superposition Principle, a consequence of the unrestricted  linearity 
of quantum theory, a microsystem can be placed in a superposition of states, e.g.,
for a magnetic atom, of ``spin up" and ``spin down" (as in the Stern-Gerlach experiment).
Then, if coupled to a measuring device, inevitably it also enters into the superposition,
e.g., of ``pointer up" and ``pointer down". Note the ``and" in that last sentence.
In life we require an ``or" when discussing mutually-exclusive events.
In his book von Neumann found, after an exhaustive
treatment of measurement scenarios, nothing available in linear QM to eliminate
one component of the superposition and leave the other. Thus, to explain measurement outcomes,
we are forced to call on external help: e.g., the ``consciousness of the observer" 
(von Neumann's solution), 
the ``multiverse" (Everett, \cite{Everett}), 
or a miracle (Pauli, \cite{BornEinsteinletters}, ,... .)

The author's proposed new term took the following form. Suppose some macroscopic object
is describable by coordinates: $x_1,x_2,...,x_N$ and let

\def\oX{\overline{X}}

\be
\oX \= \left(\,\frac{1}{N}\,\right)\,\skN\,x_k.
\ee

\ni Then the term, which might be dubbed ``WaveFunction Energy" (WFE), is:

\be
\hbox{WFE} \= w\,N^2\,D_N,
\ee

\ni where `$w$' is a (presently unknown) positive parameter and
 $D_N$, the (squared) dispersion of $\oX$, is given by:

\be
D_N \= <\psi\,|\oX^2\,|\,\psi> - \left[\,<\psi\,|\,\oX\,|\,\psi>\,\right]^2.\label{Deq}
\ee

Now the basic idea is that, in the presence of a cat, the dispersion is of large
size, and so the WFE is enormous due to the factor of $N^2$, even if the prefactor
is very small. Such a cat might have the wavefunction:

\be
\psi(x_1,x_2,...) \= \left(\,\frac{1}{\sqrt{2}}\,\right)\,\prod_{k=1}^N\,\phi_{-R}(x_k)
\+ \left(\,\frac{1}{\sqrt{2}}\,\right)\,\prod_{k=1}^N\,\phi_{R}(x_k),
\ee

\ni where $\phi_{\pm R}(x)$ is a wavefunction with narrow spatial width around the position
$\pm R$. Plugging into (\ref{Deq}),
we find that the second term nearly vanishes and the first yields:

\be
\hbox{WFE} \approx w\,N^2\,R^2.\label{cat}
\ee

Such an energy not being available in the laboratory (e.g., if $N = 10^{20}$, 
$R = 1$ m and $w = 10^{-30}$ joules/m$^2$), 
measuring devices cannot become feline,
resolving the problem. Meanwhile, applied on a micro scale of an atom or nuclei, this
term will make an infinitesimal contribution, explaining why the usual linear \Sch's equation
works so well for that regime.

In paper I, \cite{WickI}, I showed that the resulting nonlinear, nonlocal dynamics
(a) preserved the norm of the wavefunction exactly; (b) preserved overall energy
exactly (a consequence of the Hamiltonian form of the equation); and (c) explained
why classical mechanics works so well on the macro level. I also indicated
how the theory might be extended to, e.g., Dirac's 
relativistically-invariant theory.\footnote{For that extension I utilized a momentum
version of WFE obtained by substituting the momentum operator for position in $D_N$.
Which version is the better one is still undecided.} 
In subsequent publications I demonstrated that the dynamics can exhibit the phenomena
called ``chaos" and how \sdoic\ might explain why Born's {\em ad hoc} probability
hypothesis often works well, \cite{Wickchaos}, \cite{WickTDotS}.

\section{Comparing Barut's and the Author's Nonlinear, Nonlocal Terms\label{comparesection}}

For the comparison, it will be useful to make a few substitutions into the relevant
expressions. Beginning with the author's, ``double" the space variables; i.e.,
take another copy of $(x_1,x_2.,...,)$, call them $(y_1,y_2,...)$ and a doubled-arguments
wavefunction:

\be
\psi^{(2)} \= \psi(x_1,x_2,...)\,\psi(y_1,y_2,...),
\ee

I claim that in terms of this wavefunction we can write:

\be
\hbox{WFE} \= (1/2)\,w\,N^2\,<\psi^{(2)}\,|\,\left[\,\left(\,\frac{1}{N}\,\right)\,\skN\,
\left\{\,x_k - y_k\,\right\}\,\right]^2\,|\,\psi^{(2)}>.
\ee

\ni (Proof: write out the squares in the two expressions.) Now let

\def\oY{\overline{Y}}

\be
\oY \= \left(\,\frac{1}{N}\,\right)\,\skN\,y_k,
\ee

\ni and let $\rho(x)$, resp. $\rho(y)$, be the (identical) marginal densities of $\oX$ and $\oY$.
Then we have:

\be
\hbox{WFE} \= (1/2)\,w\,N^2\,\int\,dx\,\int\,dy\,\rho(x)\,\rho(y)\,\left[\,x - y\,\right]^2.
\ee

Turning to Barut's term, the second in (\ref{Baruteq}), let

\be
h^{\mu}(x) \=  
 \left(\,\frac{1}{N}\,\right)\,\skN\,j_k^{\mu}(x).
\ee

\ni and regard this quantity to be a density as above. Then we can write:

\bar
\no && \hbox{Barut's nonlinear potential energy} \= \\
\no && (1/2)\,N^2\,\int\,dx\,\int\,dy\,h^{\mu}(x)\,h_{\mu}(y)\,D(\,x - y\,).\label{BarutNLterm}\\
&&
\ear

There are several differences between the two expressions; for one, my theory was initially
proposed in a non-relativistic context, meaning that the integrals are over
space ($R^3$) and time is just a parameter.\footnote{But, for a generalization
to be relativistically-invariant in flat space-time see section 4 of \cite{WickI} and for
curved space-times
see \cite{WickSTWFS}.} Barut's set-up is relativistic and his integrals
are over space-time ($R^4$; Minkowski metric). But 
the principal distinction between the two resides in
the appearance of the Green's function in Barut's. If we restrict to the spatial
components of $x$ and $y$, and in fact

\be
D(\,x - y\,) \sim |\,x - y\,|^2,
\ee

\ni then we could conclude that, in relation to cats, the two terms have the same or similar
meaning. But the relevant Green's functions do not generally grow at infinity.
For example, the Green's function for the wave equation falls off in the spatial direction 
as $1/r$, where $r = 
\sqrt{\skN (x_k - y_k)^2}$.\footnote{\cite{FarisPDE}, equation (3.34).}

Still, there is that factor of $N^2$. So the energy required to make a cat might still
be prohibitive. But, if the factor of $R^2$ in (\ref{cat}) in my WFE is replaced using 
Barut's expression (\ref{BarutNLterm})
by $1/R$, a superposition with wavepackets of wide separation would require
less energy supplied. But, in some intermediate regime, Barut's nonlinear term
would block some cats.

\section{\TDotS\ {\em vs.} Single Events\label{warsection}}

In an article entitled ``Quantum Theory of Single Events", \cite{Barutsingleevents}, 
Barut attempted to reconcile wavefunction theory with ``single events"---without
recourse to new hypotheses. In this section I will critique his views, as I understand
them, and contrast them with my own. I am aware that this business of criticising authors no 
longer available to offer rebuttal is questionable. But I will suggest that we might have reached
a detente over the linear-\vs-nonlinear issue and its role in solving the two-slit and measurement
problems.  

I will take the two-slit conundrum first, because I think that's what Barut had in mind.
I assume the issue here is: if we turn down the light- (or electron- ) flux behind
the first screen (with the two slits) to a level where ``one photon (electron) at a time
passes through the slits" (whatever that means), and later, in each `run', see a single dot
on the second screen, then averaging over many runs of this experiment will we see a double-slit
pattern? I assume the answer is ``yes". (If ``no", then there isn't anything problematic to
discuss.)   

Barut's solution to this puzzle 
seems to be that there are two kinds of wavefunctions, which he denotes
by $\psi$ and $\Psi$: the former are narrow in space and represent ``single events",
while the latter are spread over a (perhaps) macroscopic width and represent statistical
averages of single events.  
Barut constructs solutions of various (linear) wavefunction equations, some of which
have micro spreading (e.g., no more than a Compton wavelength of an electron). But such narrow
wavepackets could spread out over time. True, potentials could be arranged that prevent
such spreading.\footnote{For example, Hans Dehmelt constructed electric and magnetic
potentials capable of trapping a single atom or electron. Once trapped, the electron's
wavefunction stays microscopic and appears to follow a Newtonian orbit.
See, e.g., \cite{WickTIB}, Chapter
15, ignoring the opening epigraph, an out-of-context \Sch\ quote.}

Returning to our two-slit experiment, suppose the flux behind the first screen
is described by a narrow (microscopic) wavefunction. Moreover, potentials have
been arranged so that, while traversing the gap between the screens, the wavepacket
cannot spread. Then surely we will see \TDotS. Over many repetitions of this experiment,
will we find a double-slit pattern? I think that impossible, as whichever slit the
wavefunction illuminated, if narrow enough it would barely, or not at all, illuminate the other.
So we would find a sum of single-slit patterns (which does not reproduce a 
double-slit pattern).

But we don't. If $\Psi$ is a double-slit wavefunction, it will not be an average of $\psi$'s.
Now suppose we drop the potentials. Then in each run the wavefunction is free to spread out
to encompass the whole of the second screen. Why doesn't the screen glow all over?
To explain a ``single event",
 one must make hypotheses about detectors (here, whatever composes the second screen)
and about the dynamics of the joint wavefunction of light (or electrons) and detectors,
perhaps along the lines of my paper entitled ``\TDotS", \cite{WickTDotS}. 
Novel, indeed nonlinear, dynamics will be necessary.

In summary: I do not think Barut found a solution for the conundrum of observing ``single events". 
Thus I charge him with
begging the question.\footnote{``Begging the question" is not synonymous with ``raising
the question". The former refers to a false mode of argument 
in which, merely by discussing an objection,
one claims to have disposed of it.}

Moving on to the MP: the issue here is ``superpositions". 
The final sentence
of his essay reads:

\begin{quote}
 
Finally, the \Sch's cat being an individual with name Kedi, with a wavefunction $\psi$(Kedi) and
not $\Psi$(Kedi), like an individual atom or an individual DNA molecule, is never in a 
superposition state!

\end{quote}

There are multiple issues with this statement. First, it is factually in error. 
In Stern and Gerlach's experiment (1922), the magnetic atoms
were without doubt in a superposition of spin-states. 
More recently, molecules 
as large as 25kDa (2000 atoms) have been placed in superpositions 
(and indeed, run through interferometers
which revealed fringes, just like in the two-slit experiment), \cite{NatPhys}. True,
this has not been accomplished with a macromolecule to my knowledge, as of the time 
I write\footnote{12 May 2025, 13:05 Pacific Standard Time.}, nor has any such molecule been
placed in a superposition of conformations (which might be biologically relevant), \cite{Arndt}. 
But it's
likely only a matter of time.

Finally, that last statement is another question-beg. One can 
readily eliminate the MP by simply declaring that microsystems are never in superpositions,
but it makes little sense mathematically, since what constitutes a ``superposition" 
depends on a choice
of a basis in the Hilbert space. Every state is a superposition of other states.
I have also encountered the claim that macrosystems (devices) all exist in ``special
states" which can evolve only into, e.g.,``pointer went up" {\em or} ``pointer went down",
but never with the {\em or} replaced by {\em and}. Merely a bigger beg.

But at one point Barut noted something that gives me hope for a commonality of views.
After remarking on how a wavepacket of microscopic width could spread to 1m,
Barut notes:

\begin{quote}

... we have considered so far linear wave equations. Now charged particles have self-energies
which add nonlinear terms to the wave equations. It is possible that these nonlinear terms ...
would imply further localization without spreading.

\end{quote}

\section{Discussion\label{discussionsection}}

Barut's inclusion of self-energy of electrons resulted in a nonlinear, nonlocal theory,
which, as was shown in section \ref{comparesection}, may be relevant for the MP.
But it is a theory of charged currents. Half of ordinary objects is made from neutral
matter (neutrons), and so the theory would have to be generalized in order to apply
to laboratory devices.

Barut's nonlinear, nonlocal term contains a Green's function that declines with distance
between the currents. This author did consider adopting such a function in postulating
his nonlinear term, but rejected the idea as adding parameters to the theory.
Barut's theory has the advantages over mine of being free of unknown parameters
and because it cannot be derided as {\em ad hoc}-ery.

The question of whether Barut's theory is ``true"\footnote{Scientific theories can not
be said to be true or false. As Thomas Kuhn maintained in his 1962 book (everyone knows his
now-iconic phrase, ``paradigm shift"), \cite{Kuhn}, theories compete with each other but are never
either proven or disproven.} or methodologically superior to QED
is not addressable by this author. (Obviously, reaching a decision
on the second part would be a big project for anyone.) As I am
first of all a mathematician, I would not want to work within a theory consisting entirely
of terms in a series known to diverge (proven for QED by Dyson in 1953). Therefore,
I would prefer Barut over Feynman {\em et al}.
 
\pagebreak

\section{Acknowledgments}
The author thanks William Faris for help with understanding Green's functions for relativistic PDEs.

\end{document}